\documentclass{elsart4-1}

\usepackage{graphicx}
\usepackage{epsfig}

\usepackage{amssymb}

\usepackage[english,francais]{babel}

\newtheorem{e-proposition}[theorem]{Proposition}

\newtheorem{e-definition}[theorem]{Definition\rm}

\renewcommand{\theequation}{\arabic{equation}}
\def\og{\leavevmode\raise.3ex\hbox{$\scriptscriptstyle\langle\!\langle$~}}
\def\fg{\leavevmode\raise.3ex\hbox{~$\!\scriptscriptstyle\,\rangle\!\rangle$}}

\begin{document}
% Select a primary header Physics or Astrophysics
% You can place after the header (classification), if you know it.

\centerline{Astrophysics}
\begin{frontmatter}


% \title{The Ankle}%\thanksref{label1}}
% \author{Name\thanksref{label2}}

\selectlanguage{english}
\title{Introduction to high-energy gamma-ray astronomy}

% If all authors are at the same address, the [label1] can be suppressed

\selectlanguage{english}
\author{Bernard Degrange \and G\'erard Fontaine}
%\ead{deligny@ipno.in2p3.fr}
%\ead{dupont@llr.fr}
%\author{Bernard Degrange and G\'erard Fontaine}
\ead{degrange@llr.in2p3.fr}
\ead{fontaine@admin.in2p3.fr}
\address{Laboratoire Leprince-Ringuet, \'Ecole polytechnique, CNRS/IN2P3,
Universit\'e Paris-Saclay, 91128 Palaiseau cedex, France}

% If your know the dates of reception, and acceptation you can put them now;
%    idem the name of the person presenting your article
% ===================================================================
%       IN CASE OF TWO AUTHORS WITH DIFFERENT ADDRESSES,
%             PLEASE USE THE FOLLOWING EXAMPLE
%\author[a]{Olivier Deligny}
%\author[b]{\and Jean Dupont}
%\ead{deligny@ipno.in2p3.fr}
%\ead{dupont@llr.fr}
%\address[a]{IPN Orsay, 15 Rue Clemenceau\\
%91406 Orsay CEDEX, France}
%\address[b]{LLR Ecole polytechnique\\
%91128 Palaiseau CEDEX,France}
% ===================================================================
\medskip
\begin{center}
{\small Published in C. R. Physique 16 (2015) 587-599}
\end{center}

\begin{abstract}
The present issue is the first of of a two-volume review devoted to gamma-ray astronomy above 100 MeV which has witnessed considerable progress
over the last 20 years. The motivations for research in this area are explained,
the follow-on articles of these two thematic issues are introduced and a brief
history of the field is given.\\
{\it To cite this article: B.~Degrange and G.~Fontaine, C. R. Physique 16 (2015) 587-599.}

\vskip 0.5\baselineskip

\selectlanguage{francais}
\noindent{\bf R\'esum\'e}
\vskip 0.5\baselineskip
\noindent
{\bf Introduction \`a l'astronomie gamma de haute \'energie.}
Le pr\'esent num\'ero est le premier de deux volumes consacr\'es \`a l'astronomie gamma de haute \'energie au-dessus de 100 MeV qui a consid\'erablement progress\'e dans les 20 derni\`eres ann\'ees.
Cet article expose les motivations \`a la base de cette recherche, pr\'esente les articles de ces deux num\'eros th\'ematiques
et fournit une br\`eve introduction historique du domaine.\\
{\it Pour citer cet article~: B.~Degrange and G.~Fontaine, C. R. Physique 16 (2015) 587-599.}

%Now keywords/mots-cls
%\keyword
\vskip 0.5\baselineskip
\noindent{\small{\it Keywords~:} Cosmic rays; Non-thermal radiation; High-energy gamma rays; Space-borne detectors; Ground-based detectors}
\vskip 0.5\baselineskip
\noindent{\small{\it Mots-cl\'es~:} Rayons cosmiques~; Rayonnement non-thermique; Rayons gamma de haute \'energie; D\'etecteurs en satellite; D\'etecteurs au sol}

\end{abstract}
\end{frontmatter}

% now the Version française abrégée, if it exists
%\selectlanguage{francais}
%\section*{Version fran\c{c}aise abr\'eg\'ee}
% Text of your Version française abrégée here

\selectlanguage{english}

%%%%%%%%%%%%%%%%%%%%%%%%%%%%%%%%%%%%%%%%%%%%%%%%%%%%%%%%%
\section{Motivations}
\label{sec:motiv}
%%%%%%%%%%%%%%%%%%%%%%%%%%%%%%%%%%%%%%%%%%%%%%%%%%%%%%%%%
\subsection{The quest for cosmic accelerators}
Present-day astronomy is still primarily concerned with the study of sources of photons and covers the electromagnetic spectrum from radio waves up to very-high-energy gamma rays. The gamma-ray domain corresponds
to photons with energies greater than 0.5~MeV and the most energetic cosmic
photons presently detected reach about 100 TeV. The present review is focused on the
high-energy part of the electromagnetic spectrum, above 100~MeV, which is related to the origin of cosmic rays. Low-energy $\gamma$-ray astronomy, which is based on
specific techniques (collimators, coded masks, Compton telescopes), and essentially addresses different questions of astrophysics (e.g. nuclear $\gamma$-ray emission lines), is not covered in the present review, except for gamma-ray bursts,
some of which have a high-energy component.

Cosmic rays, discovered in 1912 by Victor Hess \cite{Hess}, are, for the most part, high-energy protons and nuclei
whose spectrum extends over eleven orders of magnitude \cite{CRspec}, from a few times $10^9$~eV up to about $10^{20}$~eV. Their energy distribution is well described by a power law ($\propto E^{-\gamma_{cr}}$) whose exponent or ``spectral index'' $\gamma_{cr}$ is equal to 2.7 up to about $4 \times 10^{15}$~eV, then to 3 up to $4 \times 10^{18}$~eV, where $\gamma_{cr}$ is again slightly lower. The first spectral break in the $10^{15}$~eV (or PeV) region is called the ``knee''. Cosmic rays with energies below the knee are essentially originating from our Galaxy. Above the knee, their origin (galactic vs. extra-galactic) remains controversial. Therefore, identifying cosmic accelerators in which particles can reach at least the PeV energy range (or ``pevatrons'') is an important challenge today.
About 2\% of the cosmic rays are electrons and positrons with a much steeper energy spectrum \cite{CRspec}.
Unfortunately, all these particles are charged, and thus continuously deviated by turbulent magnetic fields embedded in interstellar and
intergalactic plasmas, so that their direction when they reach the Earth is almost completely uncorrelated with that of their source, except perhaps
at extreme energies where fluxes are extremely low \cite{auger}. Thus, the quest for cosmic accelerators as well as for acceleration mechanisms mostly relies on photons which propagate along a straight line. The presence of high-energy electrons and positrons can also be indirectly detected by their synchrotron
radiation from radio waves to X~rays, but the most energetic photons produced by cosmic-ray interactions with radiation fields or matter provide a complementary and more direct insight into both the accelerators and the targets (e.g. molecular clouds).
\subsection{Gamma-ray production mechanisms}
\begin{figure}
\begin{center}
\includegraphics[width=0.45\linewidth]{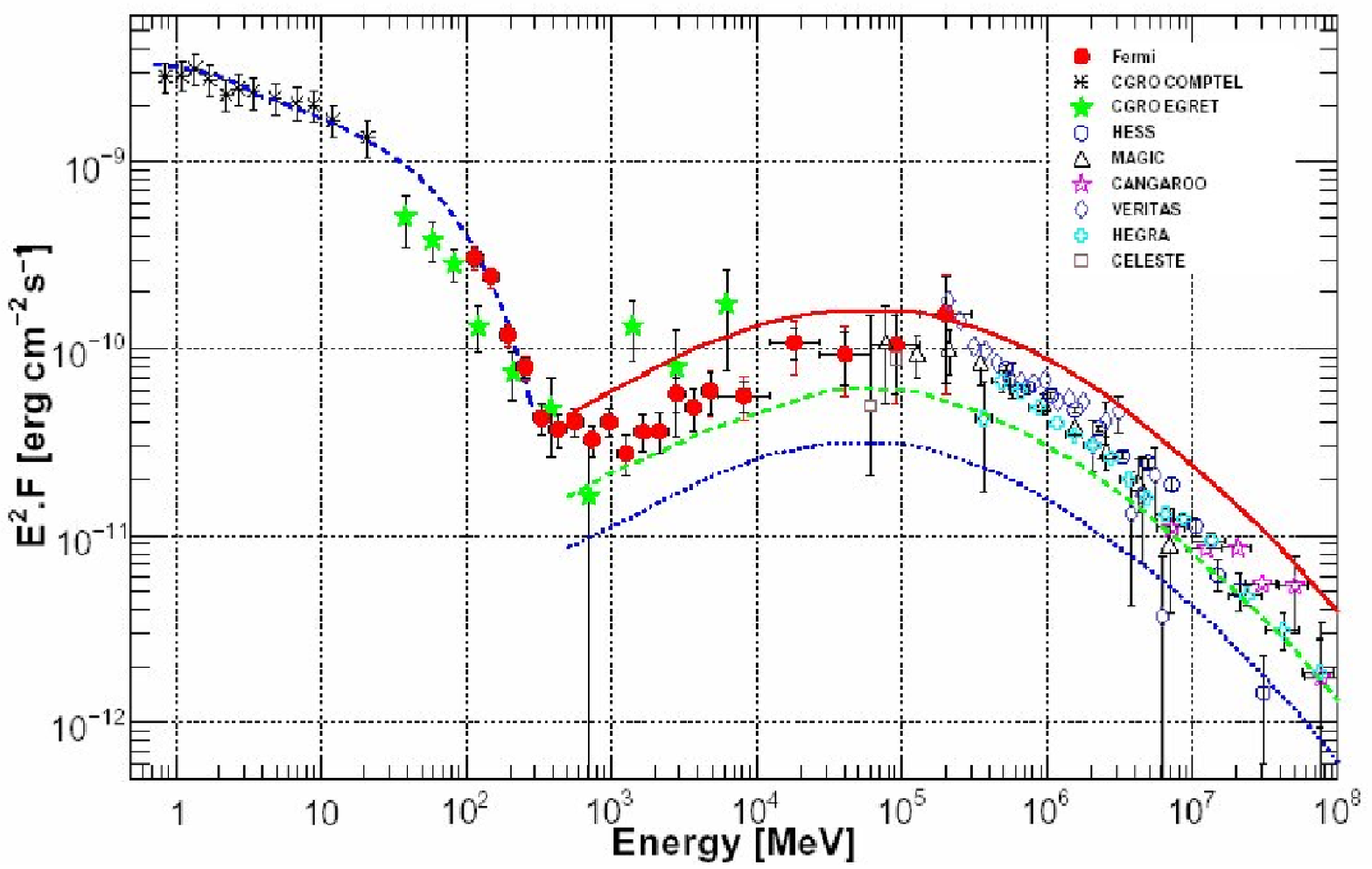}
\includegraphics[width=0.45\linewidth]{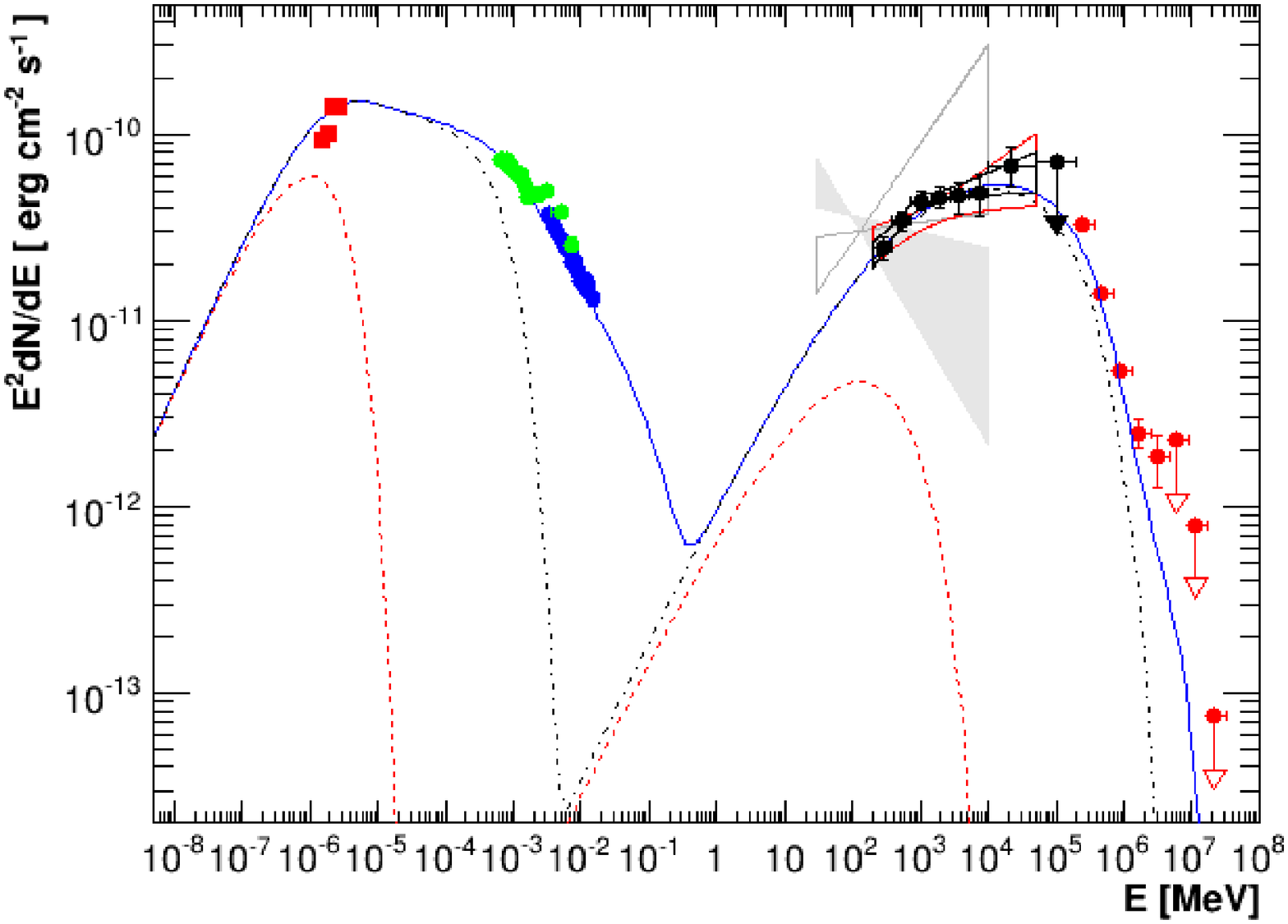}
\caption{\it Spectral energy distributions $E^2 d^3N_\gamma/(dE \, dt \, dS)$ of two different non-thermal sources emitting photons from radio waves to very-high-energy
gamma rays: on the left, a galactic source, the Crab nebula \cite{Crabsed}; on the right, an extra-galactic source, the active galactic
nucleus PKS 2155-304 \cite{PKSsed}. Note that 1 erg cm$^{-2}$ s$^{-1}$ =
$10^{-3}$ W m$^{-2}$.}
\label{fig:sed}
\end{center}
\end{figure}
Gamma rays can be produced by two different processes~:
\begin{itemize}
\item[$\bullet$] High-energy electrons and positrons interact with radiation fields. In magnetic fields, they produce synchrotron photons whose energies can at most reach
the domain of low-energy gamma rays. But they can also interact with ambient low-energy photons (from stellar or synchrotron origin) and boost them to very high
energies. This last mechanism is just the Compton effect, albeit observed from a Lorentz frame quite different from that of the electron at rest, used in nuclear
physics~; it is therefore referred to as the ``Inverse Compton effect''. The two preceding processes, induced by electrons and positrons, are said to be ``leptonic''.
\item[$\bullet$] In denser regions of the interstellar medium, high-energy protons and nuclei interact with matter through nuclear interactions, often producing
neutral mesons, mainly $\pi^0$'s, which decay into $\gamma$ rays and whose mass $m_0$ is 135~MeV/c$^2$.
These processes involving nucleons and mesons are said to be ``hadronic''. The kinematics of the decay $\pi^0 \rightarrow \gamma + \gamma$ shows that, whatever the spectrum of
the incident particle, the $\gamma$-ray energy spectrum reaches a maximum at $m_0 \, c^2/2 \approx 68$~MeV, just above the threshold of the production reaction, then decreases smoothly, according to a power law whose spectral index is close to that of the incident particle\footnote{Note that $\gamma$ rays can also undergo hadronic interactions, e.g. $\gamma$ + p $\rightarrow$ $\pi^0$ + p.}. This is the main reason why this review is limited to $\gamma$ rays above 100~MeV.
\end{itemize}
Many of the sources detected above 100 MeV also emit non-thermal photons over the whole electromagnetic spectrum. Since $\gamma$-ray differential fluxes
$d^3N_\gamma/(dE \, dt \, dS)$ decrease very rapidly with energy, it is convenient to consider the ``spectral energy distribution'' (or SED), i.e. the distribution of the
quantity\footnote{In the notation of radio-astronomy, the SED takes the form $\nu F(\nu)$ in which $\nu$ is the
photon frequency and $F(\nu)$ is the power received by unit area and frequency.}:
\begin{equation}\label{eq:sed}
E^2 \frac{d^3N_\gamma}{dE \, dt \, dS} = E \frac{d^3N_\gamma}{d \ln E \, dt \, dS}
\end{equation}
which represents the power received by unit area by unit of $\ln E$. The spectral energy distribution provides a useful representation of the non-thermal emission of
an astrophysical object from radio waves to the highest energies.
Figure \ref{fig:sed} shows the SEDs of two very different objects, the Crab nebula \cite{Crabsed}
and the active galactic nucleus PKS~2155-304 \cite{PKSsed}. For these two objects, a two-component structure is observed, the first bump being due to synchrotron radiation and the second one being
generally interpreted as due to Inverse Compton effect. In the case of PKS~2155-304 which is a variable source, the determination of the SED requires
simultaneous observations at different wavelengths in the framework of a coordinated campaign. Variability correlations may then be found between different parts of
the spectrum, thus providing important clues in modeling the object. From figure \ref{fig:sed}, it is clear that the spectra (SED or photon energy spectrum) can be approximated by power laws only in a limited energy range. A photon energy spectrum proportional to $E^{-\gamma_p}$ is said to be ``soft'' if $\gamma_p$ (photon index) is greater than 2 (which corresponds to a decreasing part of the SED), and ``hard'' in the opposite case.
In the present volume, a short introduction on acceleration and radiation
physics is given by M.~Lemoine and G.~Pelletier \cite{theory}.
\subsection{Astrophysical sources of high-energy $\gamma$ rays.}
High-energy astronomy, whose pace of development increased substantially since 1991, has now revealed several categories of
objects whose emission is dominated by non-thermal processes involving violent shocks, sometimes associated with stellar explosions or with the presence of compact objects (neutron
stars or black holes), namely\footnote{In some cases, such processes are associated to extended objects
resulting from many supernova explosions or stellar winds. Such a
``superbubble'' has recently been observed in the Large Magellanic Cloud \cite{sb_lmc}.}:
\begin{itemize}
\item[$\bullet$] Galactic sources such as pulsars, pulsar wind nebulae, supernova remnants, some particular binary systems and the centre of our Galaxy; in the present
volume, pulsars are covered by I.~Grenier and A.~Harding \cite{pulsars}, pulsar-wind nebulae and supernova remnants by J.~W.~Hewitt and 
M.~Lemoine-Goumard \cite{snr-pwn}, $\gamma$-ray emitting binary systems by G.~Dubus \cite{binaries} and 
the Galactic Center region by M.~Su and C.~van~Eldik \cite{GC}.
\item[$\bullet$] Extragalactic sources will be covered in the second volume of this review: starburst galaxies by
S.~Ohm \cite{starburst}, active galactic nuclei by C.~Dermer and B.~Giebels \cite{agn} and $\gamma$-ray bursts by F.~Piron \cite{grb}.
\end{itemize}
Extragalactic $\gamma$-ray astronomy is limited somewhat by the absorption of $\gamma$ rays on their path to the Earth by the extragalactic background light (EBL)
when the center-of-mass
energy of the $\gamma$ + photon reaction allows for the production of an $e^+ + e^-$ pair. This absorption effect, due to the infrared and optical background light,
has been observed on TeV $\gamma$ rays from active galactic nuclei. Its estimation allows the density of background photons to be measured, as explained
by D.~Horns and A.~Jacholkowska \cite{probes} in the second volume of this review. This is an important result, since direct measurements of these cosmic radiation fields are
difficult to obtain due to galactic foregrounds.
\subsection{The Universe as a laboratory of fundamental Physics}
In many respects, the Universe can be considered as a laboratory of fundamental and particle physics, in which cosmic $\gamma$-rays are useful and important tools.
High-energy $\gamma$ rays might be produced through new processes predicted by theories beyond the standard model of particle physics. In popular models of dark matter ($\Lambda$CDM models), the invisible mass
in the Universe is due to weakly interacting particles with a mass of the order of 100~GeV/c$^2$ or more (see e.g. \cite{DM}). Being their own anti-particles,
they can mutually annihilate in regions with a
high density of dark matter, thus producing high-energy photons either directly or through the decay of other particles. Gamma rays might also be produced in
micro-bursts due to the final stage of evaporation of primordial black holes through the Hawking process \cite{PBH}. 
The search for $\gamma$ rays from dark matter annihilation or primordial black holes evaporation is discussed by P. Brun and 
J. Cohen-Tanugi \cite{DM-PBH} in the second volume of this review.

New phenomena such as a possible violation of Lorentz invariance as
predicted in some theories of Quantum Gravity \cite{QG} could also affect the propagation of $\gamma$ rays from sources at cosmological distances. The observation of rapidly variable sources such as $\gamma$-ray bursts or
active galactic nuclei have yielded strong constraints on these theories. Gamma-ray propagation might also be affected by a possible quantum oscillation between a
$\gamma$ ray and an axion-like particle in the presence of external magnetic fields \cite{ALP}. As the axion-like particle is unsensitive to
background light, this oscillation would make the Universe look more transparent to $\gamma$ rays than expected; it could also produce a
modulation in the measured spectra. The search for Lorentz invariance violation and for axion-like particles is discussed 
by D. Horns and A. Jacholkowska \cite{probes} in the second volume of this review.

Although only upper bounds have been set to date on such new processes, this field of research is presently
very active.
\section{A brief history of high-energy $\gamma$-ray astronomy}
\subsection{Detecting high-energy $\gamma$ rays from space and from the ground}
The atmosphere being opaque to photons beyond the optical waveband, high-energy astrophysics required the advent of space-based experiments and started with
the discovery of the first cosmic X-ray source \cite{Xastro} by Giacconi et al. in 1962. However, as compared to soft X-ray astronomy, space-based $\gamma$-ray astronomy faces additional challenges:
\begin{itemize}
\item[$\bullet$] High-energy $\gamma$ rays cannot be focused, unlike soft X rays which can be collected by special mirrors whose effective area is much greater than
that of the detector. In the case of $\gamma$ rays, the effective detection area is practically restricted to that of the detector itself which is
necessarily limited to values of the order of 1~m$^2$ in order to fit into a launcher. Since fluxes decrease rapidly with increasing $\gamma$-ray energy,
satellite experiments can only efficiently explore the energy band below about 100~GeV, called the \textbf{high-energy} (HE) $\gamma$-ray domain as opposed to the \textbf{very-high-energy} domain (VHE), for which ground-based detectors are required.
\item[$\bullet$] Above 100~MeV, $\gamma$ rays can only be detected by their conversion into $e^+  + e^-$ pairs in matter, with the incident direction being
reconstructed from the electron tracks. Efficient detection requires converters with a short enough radiation length, but in such materials, electrons
suffer strong multiple scattering which degrades the angular resolution. This effect is however smaller as energy increases, but even in the best cases
the angular resolution is of the order of 0.15$^\circ$, whereas soft X-ray telescopes can reach values of a few arc seconds.
\item[$\bullet$] Finally, the electron, positron and subsequent $\gamma$ rays produced by bremsstrahlung are absorbed in a calorimeter which yields the total energy
of the pair with a typical resolution of 15\%.\\
\end{itemize}

The very-high-energy (VHE) domain beyond about 100~GeV requires very different techniques, operated from the ground. When a very-high-energy $\gamma$ ray enters the
atmosphere and converts at high altitude, the initial electrons and positrons generate an electromagnetic cascade: each high-energy $e^\pm$ radiates
secondary $\gamma$ rays through bremsstrahlung, which further convert into $e^+  + e^-$ pairs of lower energies. Most of those charged particles have a velocity
greater than that of light in air; on the passage of such particles, the air emits visible light through the
Cherenkov effect (see section \ref{sec-ground}). This light can be collected by special telescopes even if the electrons and positrons of the shower do not reach the ground. The
effective detection area is comparable to that of the light pool on the ground\footnote{Note that the effective detection area is not related
to the size of the mirror, the latter being relevant for the detection threshold.}, i.e. a few $10^4$~m$^2$, adapted to the very low $\gamma$-ray fluxes above 100~GeV.
Alternatively, one can detect the charged particles of a multi-TeV cascade reaching the ground in a high-altitude experiment. The main difficulty which
hampered the first ground-based experiments is that charged cosmic rays also produce such cascades in the atmosphere, which represent an enormous background compared to
genuine $\gamma$-ray-induced cascades. This difficulty was eventually overcome in 1989 with the advent of the Cherenkov imaging technique. After this, very-high-energy
$\gamma$-ray astronomy developed rapidly and reached its full maturity at the beginning of the 2000's \cite{CRAS-2000}.

In this volume, satellite detectors are reviewed by D.~J. Thompson \cite{space} and present ground-based detectors are discussed by M. de Naurois and
D.~Mazin \cite{ground}. In the second volume, future projects, both in space and on the ground, are reviewed by J. Kn\"odlseder \cite{future}.
A short history of these two different domains of $\gamma$-ray astronomy is given below.
\subsection{High-energy $\gamma$-ray astronomy in space}
\subsubsection{From OSO-3 to Fermi-LAT}
\begin{table}[htb]
\caption{\it Catalogues from space experiments in high-energy $\gamma$-ray astronomy.}
\begin{tabular}{cccccc}
Satellite & & Catalogue & Year & & Number  \\
or experiment & & & of the catalogue & & of sources \\ \hline
COS-B & & 2CG & 1981 & & 25 \\
EGRET & & 3EG & 1999 & & 271 \\
Fermi LAT & & 2FGL & 2012 & & 1873 \\
\hline
\end{tabular}
\label{tab:catal}
%\end{center}
\end{table}

\begin{figure}
\begin{center}
\includegraphics[width=0.45\linewidth]{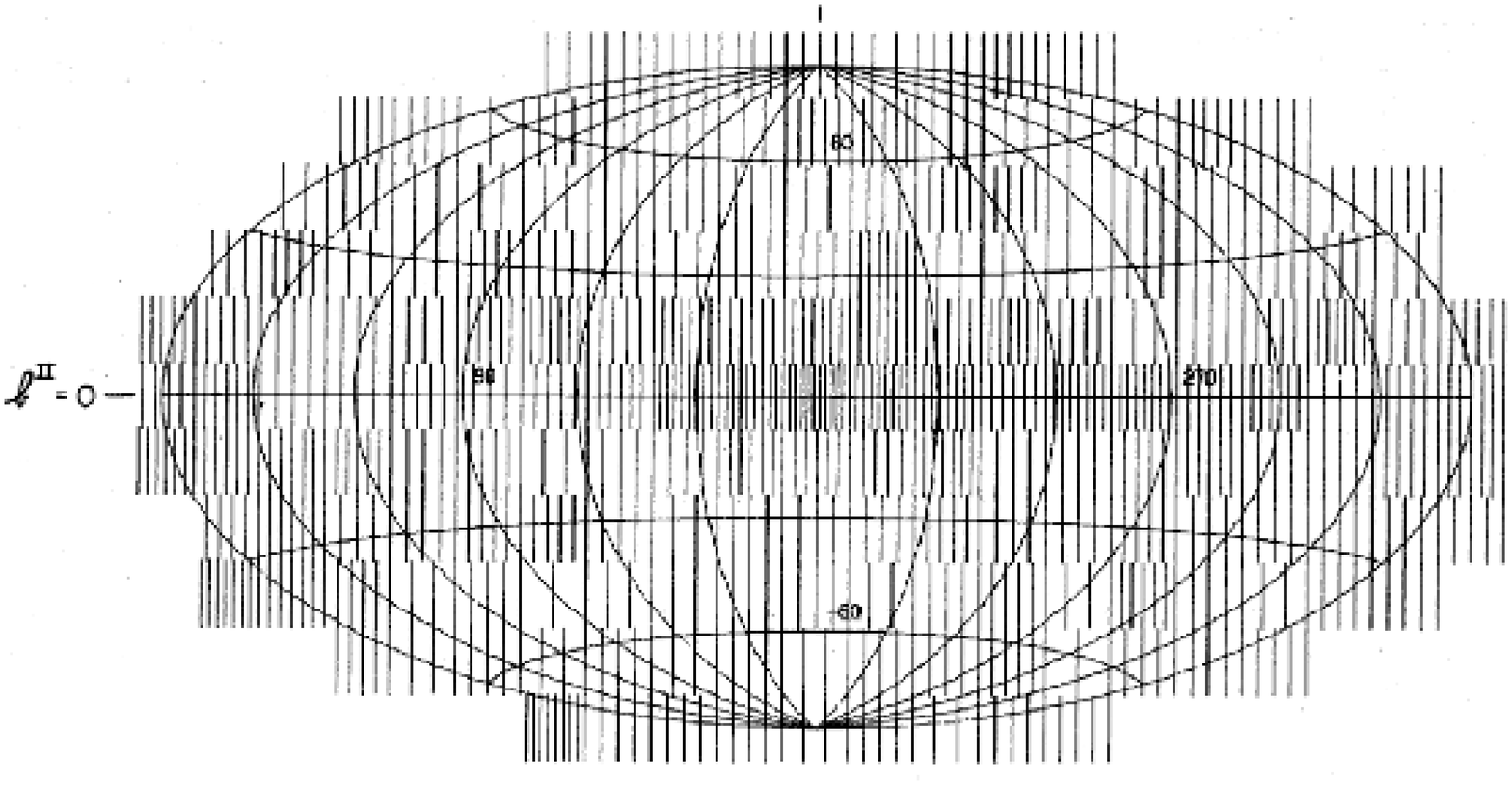}
\includegraphics[width=0.45\linewidth]{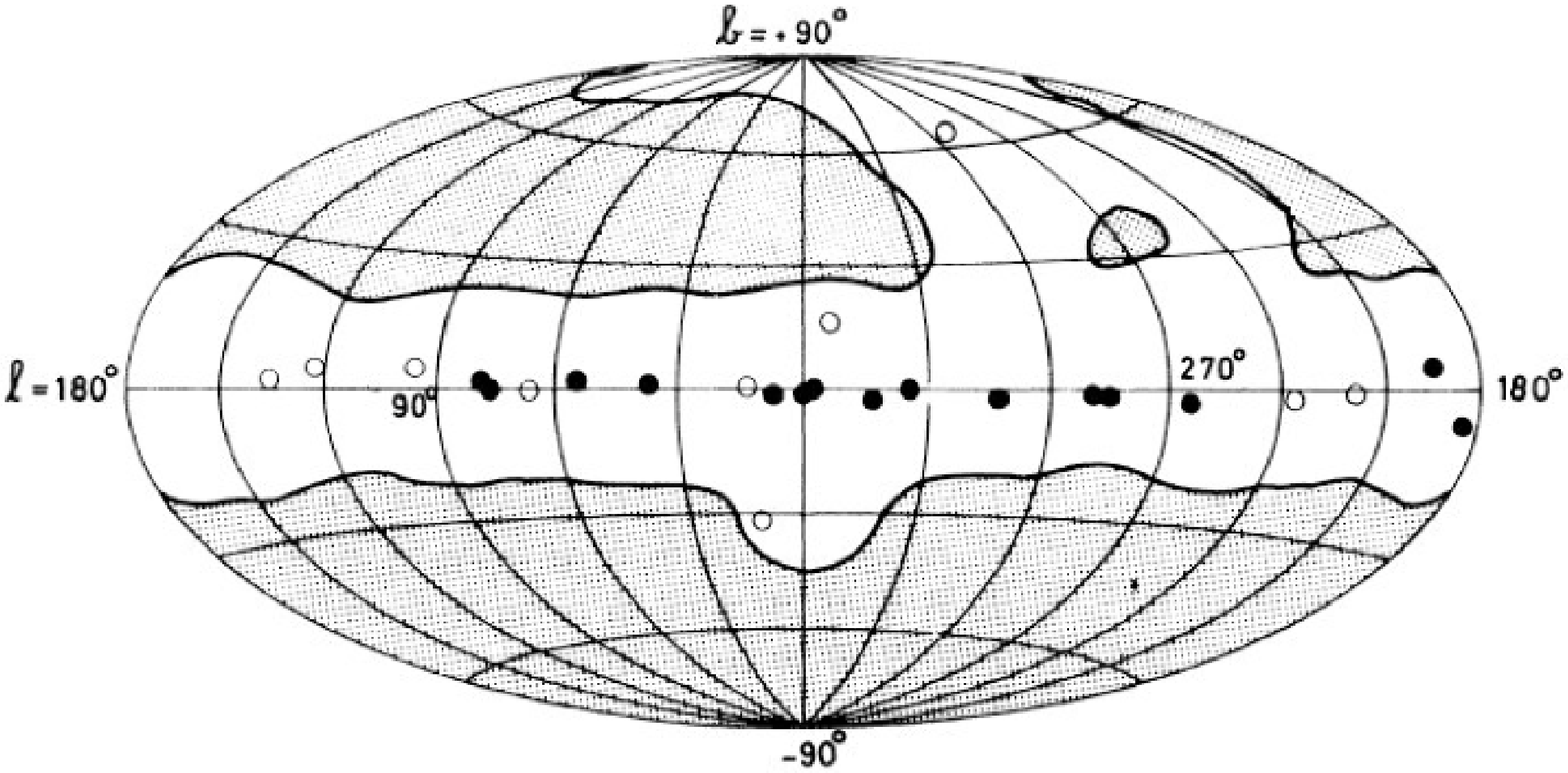}
\caption{\it First results from gamma-ray satellites: OSO-3 on the left; COS-B on the right.}
\label{fig:oso-cosb}
\end{center}
\end{figure}

\begin{figure}
\begin{center}
\includegraphics[width=0.45\linewidth]{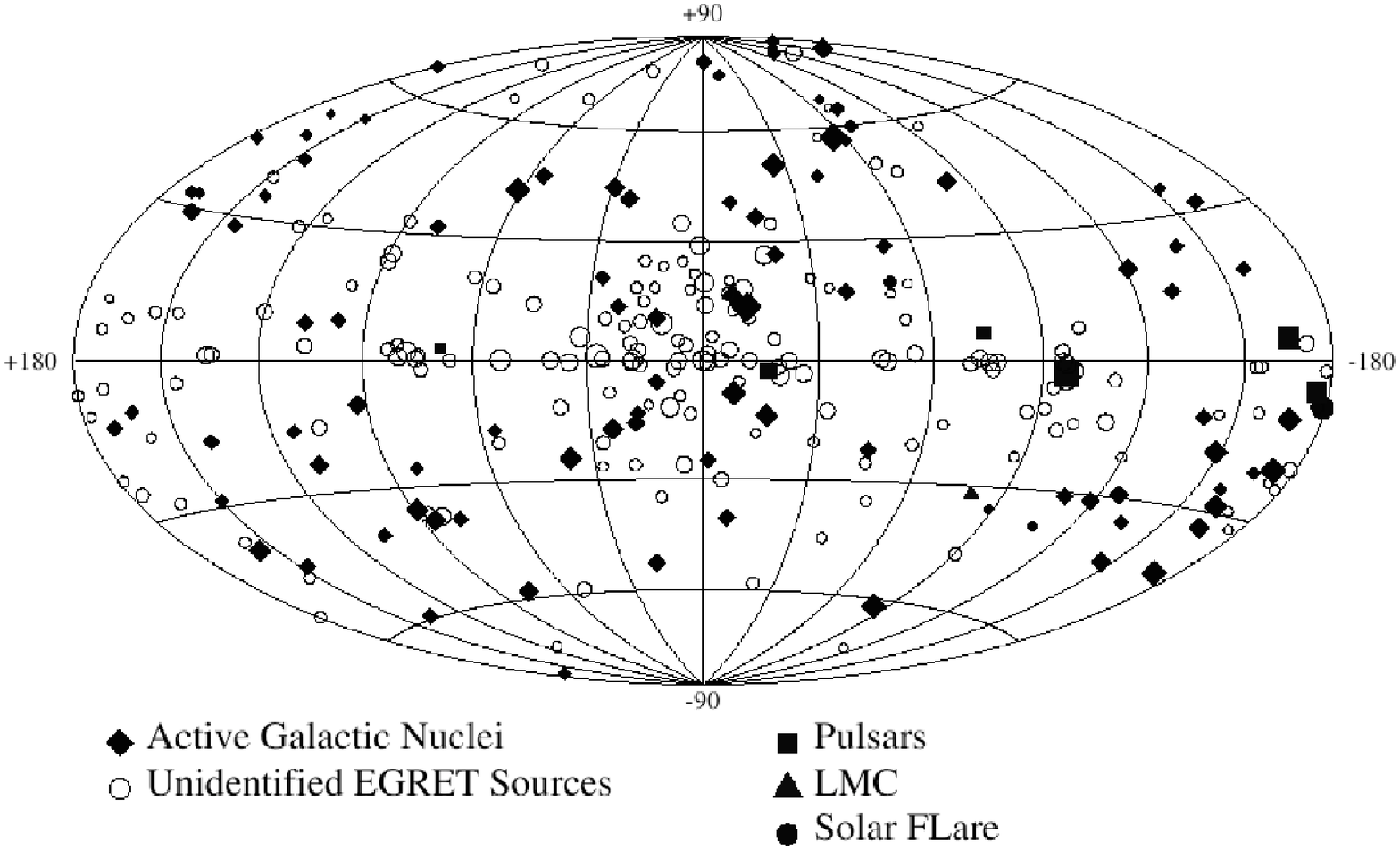}
\includegraphics[width=0.45\linewidth]{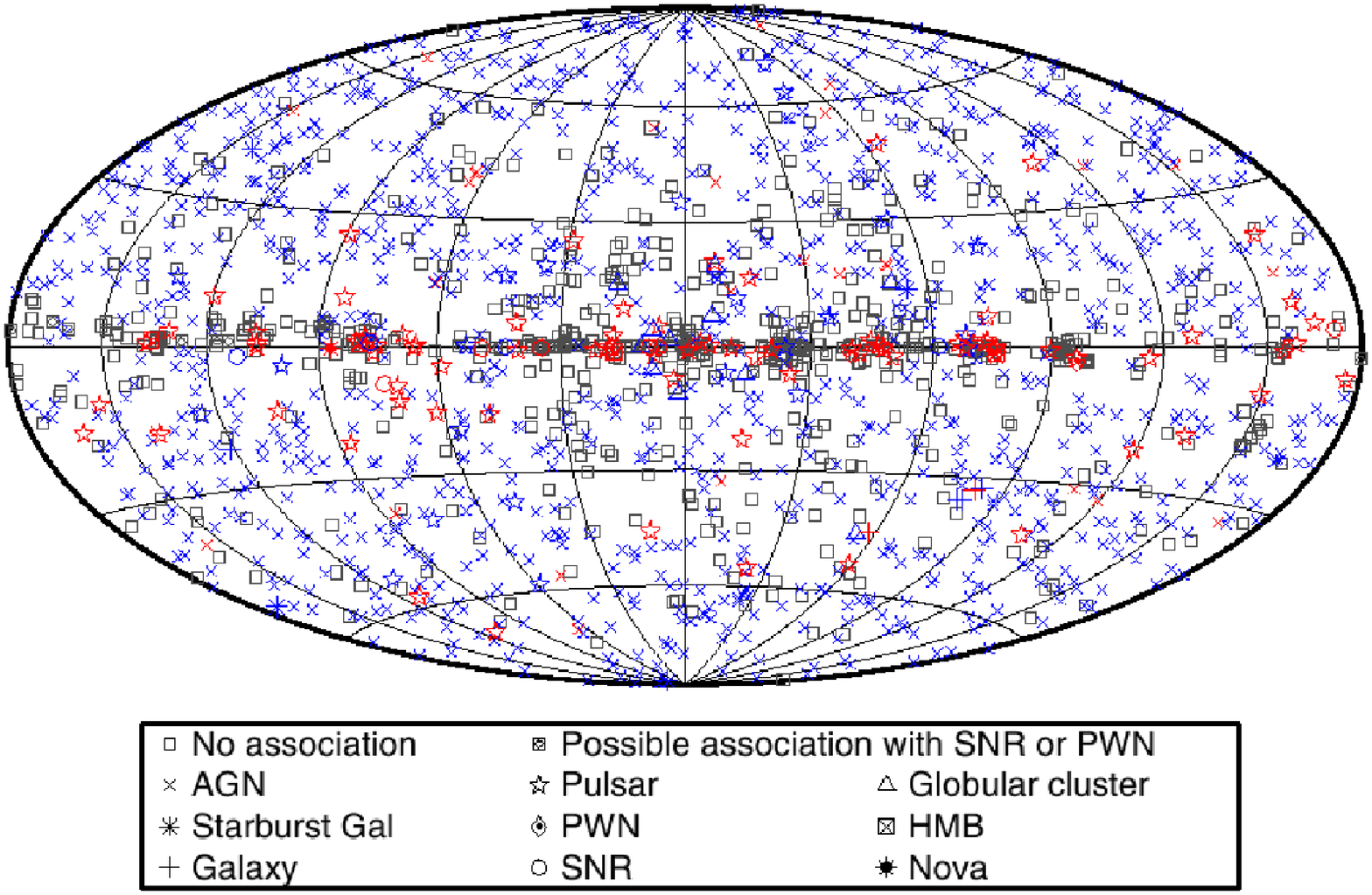}
\caption{\it Map of source locations in galactic coordinates for the third EGRET catalogue, 3EG \cite{egret_3cat} (left) and for the second Fermi-LAT catalogue \cite{2FGL} (right).}
\label{fig:egret_fermi}
\end{center}
\end{figure}

Shortly after the birth of X-ray astronomy, the first attempts to detect cosmic $\gamma$ rays with balloon-borne detectors were unsuccessful due to the high level
of secondary $\gamma$ rays produced by cosmic rays in the atmosphere. The OSO-3 satellite, flown in 1967-1968, provided the first clear evidence that the Milky Way
was a bright source of $\gamma$ rays above 50~MeV \cite{oso-3}, albeit without any imaging capability (left map in figure \ref{fig:oso-cosb}). The following
space missions in this domain were the second Small Imaging Satellite SAS-2 (1972-1973; $E > 35$~MeV) \cite{sas-2} and COS-B (1975-1982; $E>100$~MeV) \cite{cosb}.
SAS-2 revealed the diffuse emission of the Galaxy and discovered the Crab and Vela nebulae and the periodic signals from their pulsars. COS-B produced a catalogue of 25
sources (right map in figure \ref{fig:oso-cosb}), all galactic except for one, the quasar 3C273 \cite{cosb}. The next step was the Compton Gamma-ray Observatory, launched in
1991 with four gamma-ray instruments on board, among which EGRET (Energetic Gamma Ray Experiment Telescope) was devoted to high energies (20~MeV- $>$~10~GeV) \cite{egret}. The Third EGRET Catalogue
\cite{egret_3cat} revealed 271 sources (left map in figure \ref{fig:egret_fermi}) among which were many active galactic nuclei, inauguring extra-galactic
$\gamma$-ray astronomy at high energies.
The characteristics of the EGRET detector and its main results are described by D.~J.~Thompson \cite{space} in this volume. The same article explains how the
following mission, the Fermi Gamma-ray Space Telescope was conceived, taking advantage of instrumental progresses in particle physics; it describes the
Large Area Telescope (Fermi LAT) (20~MeV- $>$~300~GeV) and its performance characteristics \cite{LAT}. Launched in 2008 after a smaller
Italian mission, Astro-rivelatore Gamma a
Immagini LEggero (AGILE) \cite{agile}, it produced a particularly rich harvest of new sources (right map in figure \ref{fig:egret_fermi}), due in large part to its large effective area and huge field of view.
The Fermi Large Area Telescope Second
Source Catalogue \cite{2FGL} contains 1873 sources including various classes of objects, and a third catalogue has recently been published \cite{3FGL} 
(see \cite{space} in this volume). Table~\ref{tab:catal} summarizes the
spectacular progress of space-based $\gamma$-ray astronomy at high energies in the last 30 years.
\subsubsection{The case of gamma-ray bursts}
Gamma-ray bursts have been primarily studied below 50~MeV, but some of them have a high-energy component that is now fully accessible thanks to
the Fermi Large Area Telescope. It is therefore useful to give a short historical review on these fascinating phenomena which mostly produce
hard X~rays and low-energy $\gamma$ rays. Gamma-ray bursts
were serendipitously discovered in 1969 by the US Vela satellites monitoring the sky in low-energy $\gamma$ rays to confirm compliance with the Nuclear Test Ban
Treaty. Bursts of $\gamma$ rays coming from space were actually detected and revealed to the astronomical community only in 1973 \cite{grb_discov}. Their duration is extremely variable
(from 0.001~s to several hours) and their temporal structure quite irregular. Most of the emission takes place between a few keV and about 10~MeV, with the SED peaking around
1~MeV. The first detections could give no clue on the incident direction nor on the source distance.
The next step came from the detector BATSE (Burst and Transient Source Experiment, 20~keV-1~MeV) on board the Compton Gamma-ray Observatory. Despite a rather poor
angular resolution of about 5$^\circ$, BATSE produced a sky map of 2704 bursts \cite{batse} with a quasi-isotropic distribution, suggesting the possibility of an
extragalactic origin, which at first sight seemed unlikely due to the enormous energy required at the source. This hypothesis was nevertheless confirmed in 1997 by the
observation of GRB~970228 by the italian-dutch satellite Beppo-SAX \cite{beppo}. The burst was first detected by a wide-field camera with a 3~arc minutes angular
resolution, then the satellite was reoriented to localize the source within 50 arc seconds with narrow-field X-ray telescopes. This allowed several ground-based telescopes
as well as the Hubble space telescope to associate the source with a fading optical transient emission about 20 hours later; the extragalactic
nature of the source was then established by localizing this ``afterglow'' within a host galaxy whose redshift was found to be 0.835. Further space
missions (HETE-2, INTEGRAL, Swift) strongly contributed to the study of $\gamma$-ray bursts at low energies, whereas the corresponding optical and radio afterglows
were subsequently followed from the ground. These observations confirmed that sources were extragalactic, sometimes located at
cosmological distances, e.g. that of the burst GRB~090423 detected by the Swift satellite, whose optical counterpart yielded a redshift of 8.26 \cite{grbz}.
These phenomena are currently interpreted as strongly beamed emission originating from collapsing massive stars or merging compact objects.

Until the advent of the Fermi mission, little was known about the emission of $\gamma$-ray bursts at high energies. EGRET was not
adapted to detecting them, due to the large dead time of 0.1~s of its spark chamber; it nevertheless recorded some,
such as GRB~940217 in which an 18~GeV $\gamma$ ray was detected more than 90 minutes after the burst began \cite{grb94}. On board the Fermi Gamma-ray Space Telescope,
$\gamma$-ray bursts are detected by the Gamma-ray Burst Monitor (GBM) \cite{gbm} from 8~keV to 40~MeV and by the Fermi LAT from 20~MeV to 300~GeV.
Unlike EGRET, the LAT has a short dead time of 26~$\mu$s, providing efficient detection of the high-energy component of $\gamma$-ray bursts triggered
by GBM. After 3 years of observation, the LAT observed 35 bursts at energies greater than 20~MeV \cite{grb_LAT}; they are among the brightest ones
detected by GBM. An exceptionally bright burst, GRB~130427A, showed a high-energy component temporally distinct from that detected by GBM and lasting
almost 20 hours, with a 95~GeV $\gamma$ ray detected a few minutes after the burst began. These very recent data, which cannot be explained without introducing an
additional component in the GRB spectra, are discussed in this volume by F.~Piron \cite{grb}.

\subsection{Very-high-energy $\gamma$-ray astronomy from the ground}
\label{sec-ground}
What is now known as Cherenkov radiation was studied experimentally by P.A. Cherenkov \cite{Cherenkov 1934} in 1934 while conducting experiments on beta particles travelling through water, and it wasn't until 1937 that this emission was explained theoretically by I. Frank and I. Tamm \cite{Frank 1937}. Most of the work on Cherenkov radiation at that time was done with charged particles passing through solid and liquid media, but in 1947 P.M.S. Blackett \cite{Blackett 1948} pointed out that the effect should exist in gases and suggested that Cherenkov radiation produced by cosmic ray particles traversing the atmosphere could contribute to the light intensity of the night-sky.
Such pulsed visible light emission induced by cosmic particles in the atmosphere was indeed detected for the first time in 1952 by W. Galbraith and J.V. Jelley \cite{Galbraith 1953} in association with large cosmic ray air showers, and it was later shown to be genuine Cherenkov radiation by the same team in a series of measurement done at the Pic du Midi (France) during the summer 1953. This led A. Chudakov \cite{Chudakov 1958} to implement the first Atmospheric Cherenkov Telescope (ACT) as a way to perform a calorimetric measurement of the energy of extensive air showers (EAS) in complement of a ground counter array in Pamir at 3860 m asl.

The switch of interest from cosmic rays (CR) to very high energy (VHE) cosmic gamma-rays can be traced back to a presentation by G. Cocconi
\cite{Cocconi 1960} at the 1959 International Cosmic Ray Conference (ICRC) in Moscow in which he predicted that the Crab Nebula should be a strong source of $\gamma$ rays at TeV energies and that a high angular resolution could make possible to separate them from the isotropic CR background.
Even if Cocconi's model overestimated the flux by a factor 1000, and the proposed detector was not really appropriate, this contribution stimulated further work to use Cherenkov radiation as suggested by G.T. Zatsepin and A.E. Chudakov \cite{Zatsepin 1961}, soon followed by the construction in Crimea of the first ACT designed for gamma-ray observations in the early 1960s.
But the charged cosmic-ray background was too huge for this first generation of ACTs: they could not discriminate between electromagnetic and hadronic
showers, and hence could only rely on a subtraction method (``counts on source'' minus ``counts off source''), so they were not sensitive enough to detect successfully cosmic gamma-rays, even from the strongest sources.
\subsubsection{The long way towards a first firm detection}
A landmark was set by the construction, completed in 1968 on Mount Hopkins (Whipple observatory) in Arizona, of a 10~m diameter ACT, by far the largest reflector dedicated to this field of research for many years. Again the difficulty to overcome the huge background was only slowly overcome after a long series of detector improvements and analysis method developments.
The instrumentation of the focal plane of this telescope went through a series of steps, from a single phototube, through a set of 2 tubes in coincidence in 1972 for the so-called double beam technique, then a set of 3 tubes with a guard ring of 7 additional tubes in 1976, and after a stop of funding from 1978 to 1982 reached the 37 pixel imaging camera proposed by Trevor Weekes in 1981 \cite{Weekes 1981} and completed in 1983.
In the meanwhile, Monte Carlo computer simulations of the shower development and Cherenkov light emission and detection was more and more used to help understand the response of the detectors.
Together with the focal plane instrumental progress, it allowed the development of image analysis methods, such as that proposed by A.M. Hillas in 1985 \cite{Hillas 1985} (and widely known as the Hillas parameter method), that greatly improved the background rejection.
This was decisive for obtaining the first successful detection of the gamma-ray emission above 0.7 TeV from the Crab nebula in 1989 by the Whipple collaboration \cite{Weekes 1989}, 37~years after the initial Cherenkov light pulse observation by W. Galbraith and J.V. Jelley!
\subsubsection{The flourishing of new detectors and the confirmation of the Crab signal}
During this period, there were active experiments in Ireland, the United States, the U.S.S.R., Australia and India, but many results only had a
weak 3-4 $\sigma$ (standard deviation) statistical significance with systematics not always well controlled.
A discussion took place in 1985 at the ICRC conference, which led to an informal agreement for positive detection to require at least 5 $\sigma$ of statistical significance including the effect of trial factors.

Efforts in this field have been strongly stimulated by the claim in 1983 by two experiments (the Kiel and the Haverah Park arrays) of a possible
detection of $10^{15}$~eV $\gamma$ rays from Cygnus X-3 with an abnormal muon content in the showers that could have been interpreted as a possible new physics.
This result was highly controversial, see \cite{Bonnet-Bidaud 1988} for a full discussion, but it created some interest in the HEP community and triggered a move of new teams to this field that brought in their own expertise and technology.

Other Cherenkov techniques, developped in France and based on a sampling of the Cherenkov wave front with many telescopes spread on the ground also proved to be efficient in rejecting the night sky and hadronic background, and they were able to confirm the Crab nebula emission at VHE energies : ASGAT \cite{Goret 1993} in 1992 and THEMISTOCLE \cite{Baillon 1993} in 1993.
This wavefront sampling technique was subsequently extended to lower energies (around or below 100 GeV) by using fields of large heliostats built for solar energy plants (so called solar farms) with the addition of a secondary optics such as in experiments like CELESTE \cite{Pare 2002} \cite{de Naurois 2002}, STACEE \cite{Gingrich 2005}, GRAAL \cite{Arqueros 2002} and Solar Two \cite{Tumer 1999} which were able to detect the strongest sources at the level of the Crab nebula flux, but suffered from a limited field of view hampering their background rejection.
\subsubsection{Further developments of the imaging technique}
In the years following the successful detection of the Crab nebula, most of the progress came from further developments in the technique of imaging atmospheric Cherenkov telescopes (IACT) with a progressive increase of the granularity of image detectors.
For instance the camera at the focus of the Whipple telescope first upgrade from 37 to 109 pixels allowed the detection a second VHE source which turned
out to be extragalactic : an AGN (active galactic nucleus) Mrk 421 \cite{Punch 1992}.
It was soon followed by a series of further upgrades to 151 pixels in 1996, 331 pixels in 1997 and 490 pixels in 1999.

Building upon the development of the imaging technique, and following a dedicated workshop organized in 1992 by P. Fleury and G. Vacanti in Palaiseau, several new IACT came also online worldwide in the 1990’s such as the CANGAROO I telescope \cite{Hara 1993} operated from 1992 to 1998 with a 220 pixel camera; or the MARK VI detector \cite{Armstrong 1999} operated from 1995 to 2000 with a central imaging (109 pixels) telescope complemented by two additional simpler telescopes for triggering purpose.

During this period two important advances were achieved:
\begin{itemize}
\item[$\bullet$] The power of stereoscopic observation of air showers was demonstrated by the HEGRA collaboration operating a set of 5 telescopes \cite{Daum 1997} from 1996 to 2002, each with a 271 pixel camera, allowing a better assessment of the shower geometry, and hence an improved rejection of the hadronic background and yielding an increase in sensitivity of about a factor 10 compared to a single telescope of the same size.
\item[$\bullet$] The importance of having a very fast and fine grained camera associated with a more sophisticated analysis was demonstrated by the CAT collaboration \cite{Barrau 1998} \cite{Le Bohec 1998} using, also from 1996 to 2002, a telescope with a 546 pixel camera and a 12 ns integration time, allowing an energy threshold comparable to that of the Whipple telescope to be reached with a much smaller mirror (16~m$^2$), compared to that of Whipple (60~m$^2$).
\end{itemize}
\subsubsection{Present major IACT systems}

%%%%%%%%%%%%%%%%%%%%%%
% Insert here Table 2
% Table 2 : Main characteristics of present major IACT systems.
%%%%%%%%%%%%%%%%%%%%%%

\begin{figure}[hbt]
\begin{center}
%\begin{wrapfigure}{r}{6cm}
\includegraphics[width=0.5\linewidth]{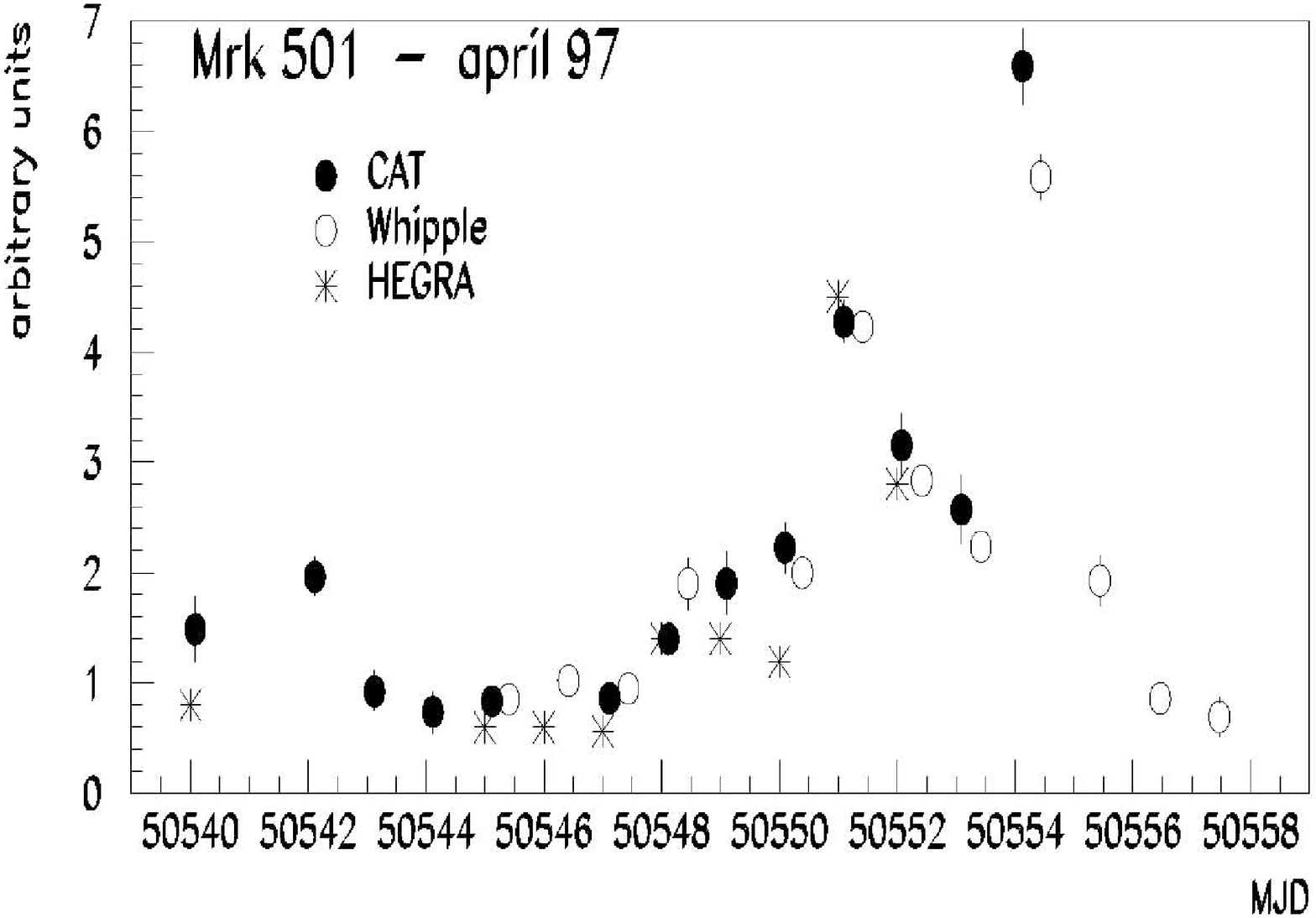}
\caption{\it April 1997 nightly gamma-ray rate from the active galactic nucleus Mrk~501, as seen by three experiments: CAT, Whipple observatory and HEGRA.}
\label{fig:mkn97}
%\end{wrapfigure}
\end{center}
\end{figure}

\begin{figure}[h,b,t]
\begin{center}
\includegraphics[width=0.9\linewidth]{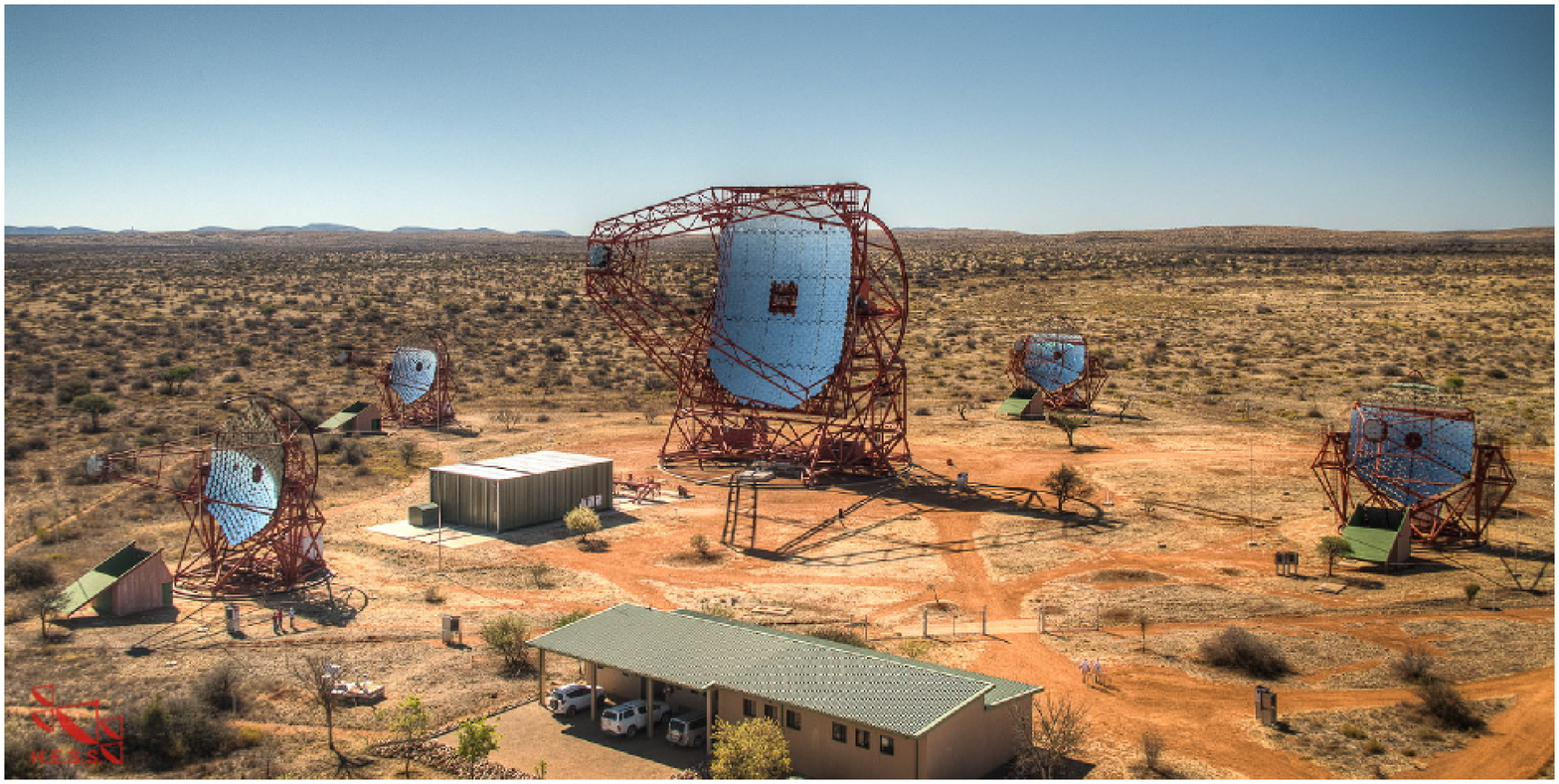}
\includegraphics[width=0.33\linewidth]{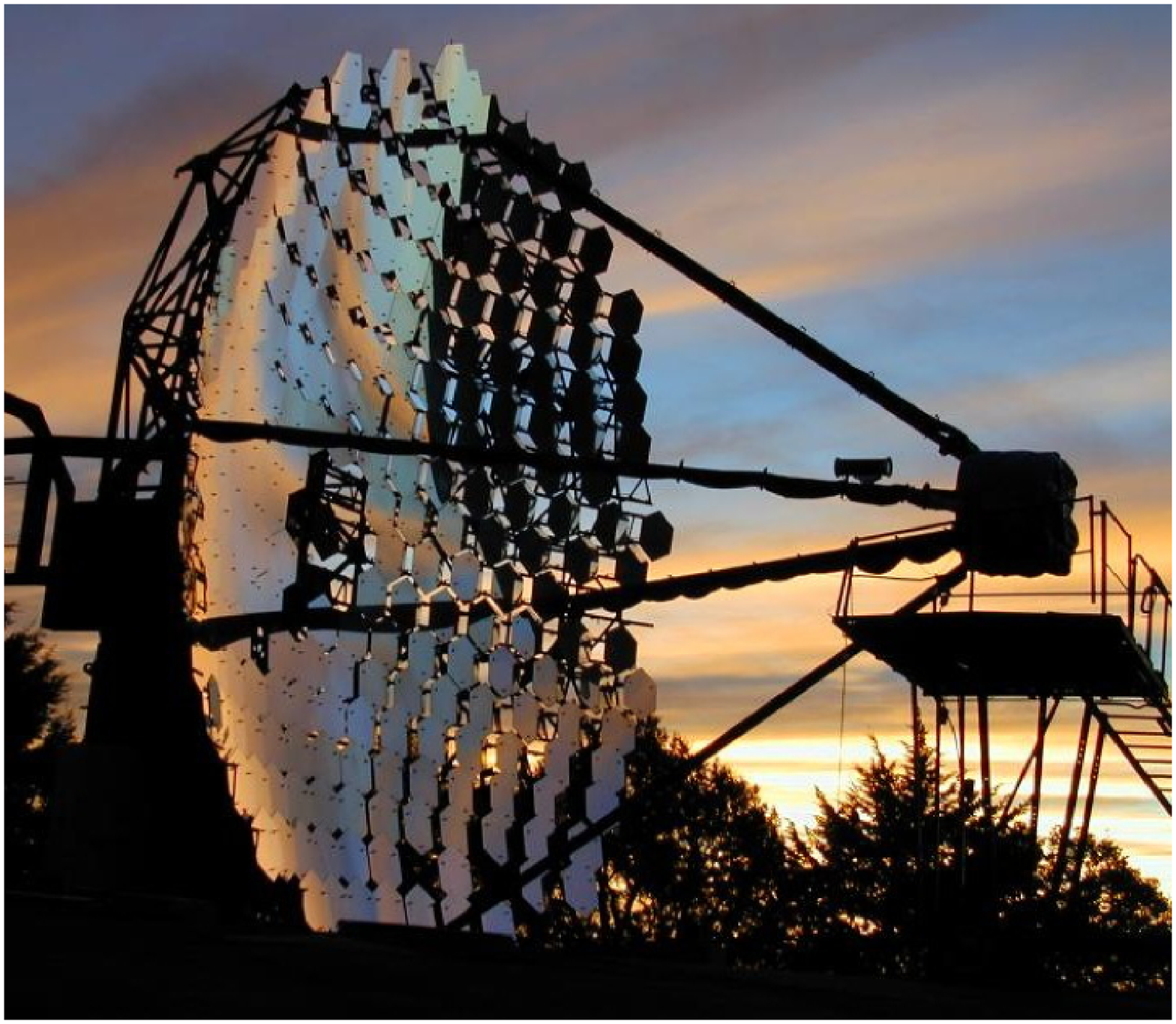}
\includegraphics[width=0.32\linewidth]{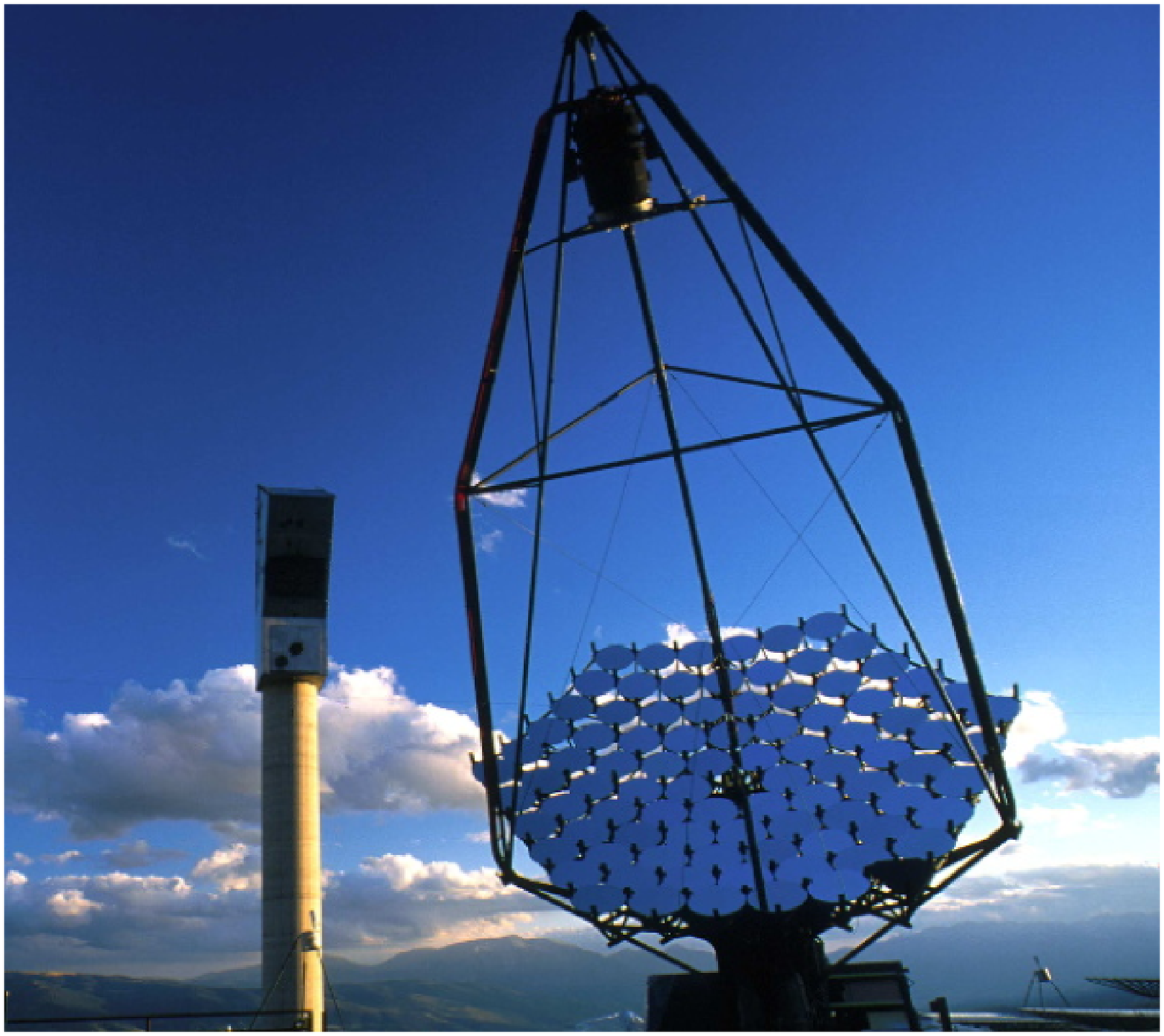}
\includegraphics[width=0.32\linewidth]{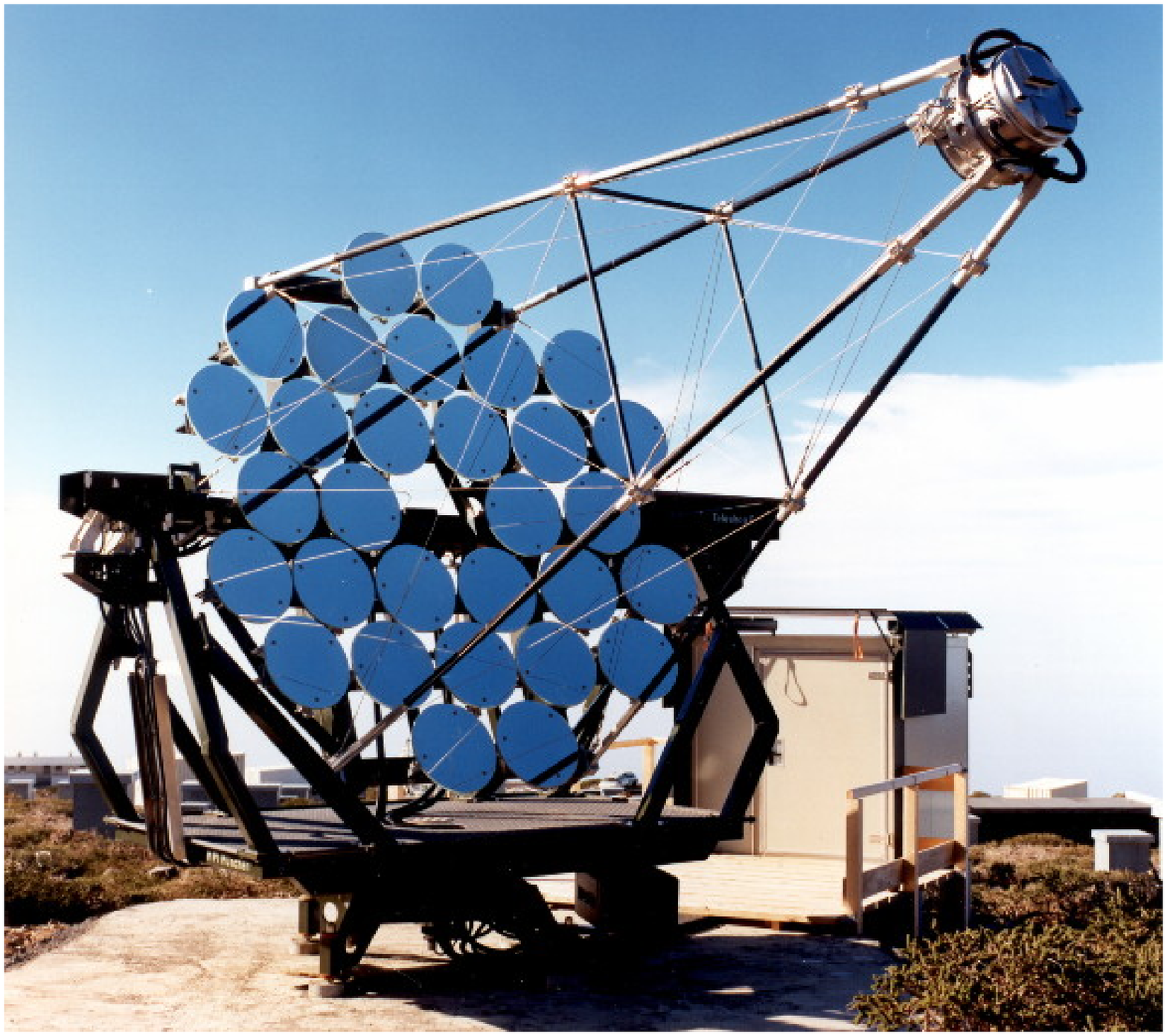}
\caption{\it Illustration of the combination in H.E.S.S. (upper panel) of large dishes as in Whipple (lower left), fast and fine grained cameras as in CAT (lower centre) and stereoscopic observation as in HEGRA (lower right)]. Image credits: Christian F\"ohr, MPIK - Whipple TBD - CNRS Phototh\`eque/IN2P3 Fran\c{c}ois Toussenel - HEGRA TBD}
\label{fig:hwch}
\end{center}
\end{figure}

Another milestone was reached in April 1997 when 3 major experiments (CAT, HEGRA and Whipple observatory) detected the same flaring episode of the AGN Mrk 501. 
It was the first simultaneous detection of a flaring source and the coherent light curve, assembled from these observations (see figure \ref{fig:mkn97}) and taking benefit 
of their spread in longitude, was presented \cite{Djannati 1997} at the Kruger Park workshop (the 5th in the series initiated at Palaiseau in 1992).
This new and exciting result brought much confidence and credit to the field and induced a strong push to build new facilities combining the
best of these three detectors: large dishes to lower the energy threshold, fast and fine grained cameras to get the best from image analysis and
stereoscopic observation with several telescopes as illustrated in Fig.~\ref{fig:hwch}.

%%%%%%%%%%%%%%%%%%%%%%
% Reserve here some space for a square figure about 16 x 16 cm
% Sources of images:
% HESS adapted from Christian F\"ohr, MPIK:
% http://www.mpi-hd.mpg.de/hfm/HESS/pages/home/Webgalleries/HESS2_telescope/images/120930_Namibia_Foehr_2777_79_81_83_85.jpg
% Whipple:
% ???
% CAT from CNRS Phototh\`eque/IN2P3 TOUSSENEL François R\'ef\'erence 2001N00905:
% http://phototheque.cnrs.fr/s?ccabidiuocmijspg
% HEGRA
%
% [Figure 4.  Illustration of the combination in H.E.S.S. of large dishes as in Whipple, fast and fine grained cameras as in CAT and

%%%%%%%%%%%%%%%%%%%%%%

The first of these new projects was the evolution, starting in 1997, towards Cangaroo II and then to Cangaroo III \cite{Kubo 2004}, which led to the operation from 2003 to 2011 of a 4 telescope array (3 tel. only beyond 2004), of 10~m IACTs equipped with a 427/552 pixel camera.
But due to various factors, among which the technical choices for phototubes and the mirror technology, its sensitivity was quickly outperformed by the present major IACT systems : H.E.S.S. \cite{Stegmann 2012 & Bolmont 2014}, MAGIC \cite{Aleksic 2012} and VERITAS \cite{Kieda 2013}. These large VHE gamma-ray observatories are presented with more details in a separate paper of this volume \cite{ground}, and their
main characteristics are summarized in Table~\ref{tab:iact}.
\begin{table}[thb]
\begin{center}
\caption{\it Main characteristics of present major IACT systems.}
\begin{tabular}{|c||c|c|c|c|c|}\hline
System  & Location & ~Number of~ & ~Mirror area~ &  ~Number of~ & Date of \\
        &          & telescopes &            & pixels & first light \\ \hline \hline
	&          &  4     & $4 \times 108$~m$^2$ & $4 \times 960$ & Dec. 2003 \\
H.E.S.S. & Namibia &        &                      &                &          \\
        &          &  1     & +~614~m$^2$          &  +~2048        & July 2012 \\ \hline
	& La Palma &        &                      &                & ~2004 (1 tel.)~ \\
MAGIC   & Canary   &  2     & $2 \times 236$~m$^2$ & $2 \times 576$ &                \\
        & Islands  &        &                      &                & ~2009 (2 tel.)~ \\ \hline
        & ~near Tucson~ &     &                      &                 &                \\
VERITAS &           & 4     & $4 \times 110$~m$^2$ & $4 \times 499$ & Jan. 2007 \\
        & Arizona  &        &                      &                 &      \\
\hline           		
\end{tabular}
\label{tab:iact}
\end{center}
\end{table}

H.E.S.S. (High Energy Stereoscopic System) has been designed as a general purpose detector for observing the southern sky (where most of our galaxy is located) with an unprecedented sensitivity. It started observations in 2004 and was awarded the Descartes Prize of the European Commission in 2006 - the highest recognition for collaborative research - and the prestigious Rossi Prize of the American Astronomical Society in 2010. Its initial energy threshold (100 GeV for the four telescope system) has recently been improved (below 50 GeV) by the addition of a fifth larger telescope.

The MAGIC collaboration started with a single large telescope optimized for the low energy observation of transient phenomena and the
study of distant AGNs. It was later complemented by a second identical telescope to allow stereoscopic observations and benefit of the
associated background reduction that allowed the detection of the pulsed $\gamma$-ray emission of the Crab in the 50-400 GeV energy range.

Finally, VERITAS (Very Energetic Radiation Imaging Telescope Array System) came online in 2007.
It has recently (2012) been upgraded to include a better trigger system, higher quantum efficiency phototubes, and high speed networking, with the
result of an improved $\gamma$-ray sensitivity and an energy threshold reduced by 30\%.
Since 2007, it has detected more than 20 extra-galactic objects, and in the recent years, its focus has shifted from discovery of new targets to long-term monitoring of known sources.

The wealth of the scientific harvest of these IACT has pushed all groups to unite in the preparation of the world-wide CTA (Cherenkov Telescope Array) project \cite{Acharya 2013} which aims at expanding the energy coverage, improving the angular resolution and increasing the sensitivity by a factor 10.
\subsubsection{Ground-based detectors with a wide field-of-view}
Another technique, based on revisiting surface arrays must also be reported here: scintillation counters are now replaced by big water tanks in which a large fraction of the shower particles that reach the (high altitude) ground are detected through their Cherenkov light emission in water. This allows dense arrays to be built, reaching a high efficiency of particle detection over a large area, and now makes it possible to detect astrophysical $\gamma$-ray sources. This water Cherenkov technique gave its first source detections with the MILAGRO \cite{Abdo 2006} array (2000-2008). The present major detector of this type is the HAWC (High Altitude Water Cherenkov) $\gamma$-ray observatory \cite{Abeysekara 2013} whose construction has just been completed using 55 kilotons of water distributed over 300 tanks at an altitude of 4100 m a.s.l.
Its energy threshold will be higher than IACT's and its hadron rejection and angular resolution will not reach the IACT level, but
HAWC will observe continuously (while IACT have a maximum of 20\% of duty cycle), have a much wider field of view (though its effective energy threshold increases rapidly with zenith angle) and offer a good stability and ease of operation. So, HAWC will be very complementary to IACT and will notably allow a full sky survey at TeV energies, the detection of unexpected transients for providing alerts to pointed instruments, and the study of large extended sources.

\subsubsection{Presently detected VHE $\gamma$-ray sources: a rich catalogue}
\begin{figure}
\begin{center}
\includegraphics[width=0.40\linewidth]{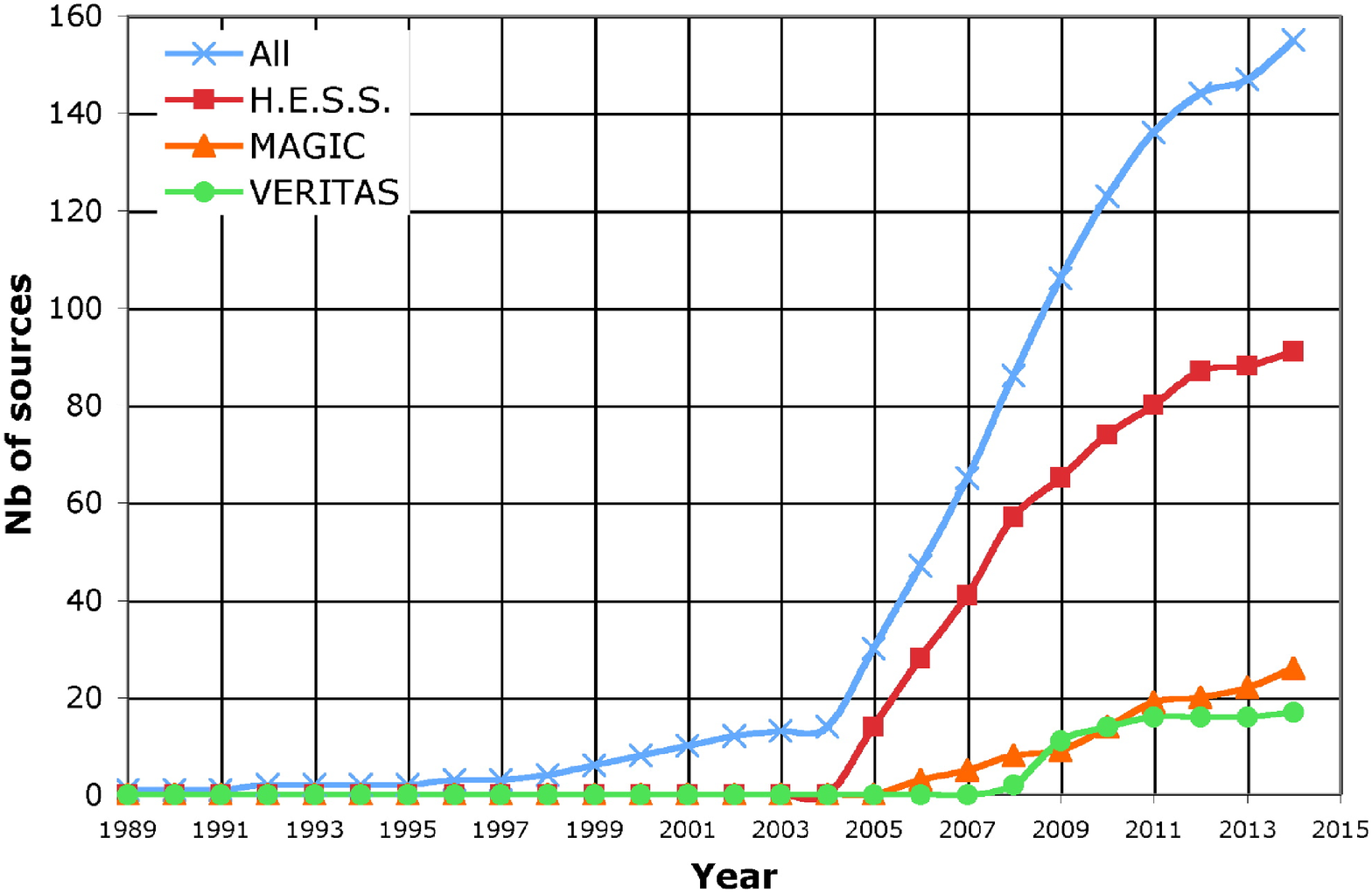}
\includegraphics[width=0.50\linewidth]{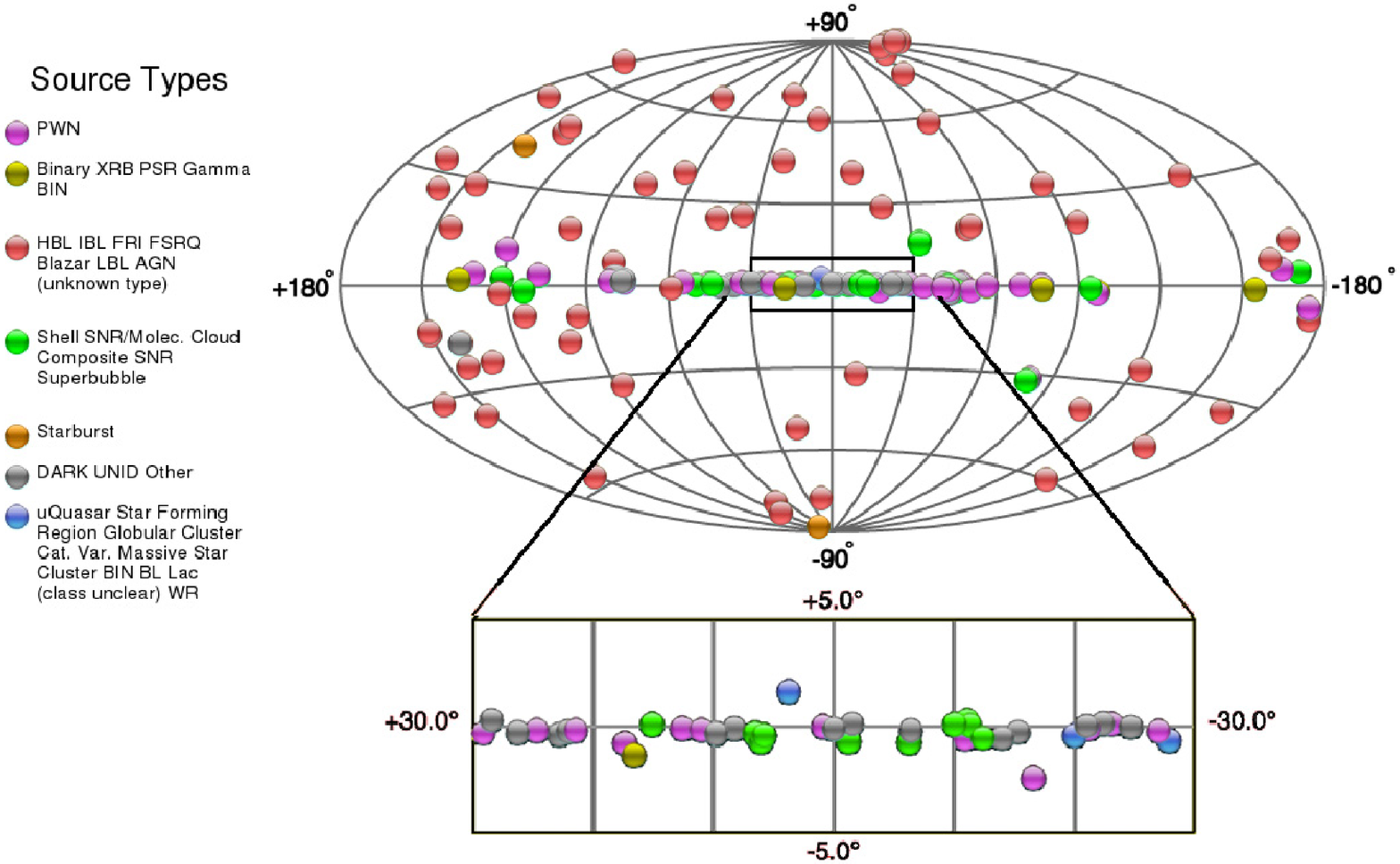}
\caption{\it (a) Evolution over time (year of announcement) of the number of VHE gamma-ray sources (TeVCat \cite{TeVCat}), with the contributions
from H.E.S.S., MAGIC and VERITAS. (b) Sky map of the present 155 VHE gamma-ray sources (TeVCat \cite{TeVCat}) in galactic coordinates (Hammer projection)
with a zoom on the Galactic Centre area.}
\label{fig:vhe_sources}
\end{center}
\end{figure}

All the VHE gamma-ray sources and the associated publications are registered in an online catalogue, TeVCat \cite{TeVCat}, from which sky maps and characteristic tables can easily be extracted.
This database shows that 155 highly significant sources have now been published in referred journals (or recently announced).
Fig.~\ref{fig:vhe_sources}a displays how this number has grown over time, from the first discovery in 1989 to the end of 2014. It clearly exhibits a slow evolution until 1996 (only 3 sources by that time), a more visible slope over [1996-2004] due to camera upgrades or new telescopes coming online, and a very fast rise from 2005 onwards when H.E.S.S. started operations, soon followed by MAGIC and VERITAS and bringing the source count to its present level of 155.

%%%%%%%%%%%%%%%%%%%%%%
% Insert here Fig. 5
% Figure 5.  Evolution over time (year of announcement) of the number of VHE gamma-ray sources (TeVCat \cite{TeVCat}), with the contributions from H.E.S.S., MAGIC and VERITAS.
%%%%%%%%%%%%%%%%%%%%%%

A sky map of these VHE gamma-ray sources in galactic coordinates is shown in Fig.~{\ref{fig:vhe_sources}b. The contribution of the galactic sources is concentrated close to the horizontal axis and most of the sources situated away from this axis are extragalactic active galactic nuclei. The physics properties of these various objects are discussed in the other papers of this topical issue of Comptes Rendus Physique.

It was not possible to cover all experiments in this short historical introduction, and a more comprehensive coverage of VHE gamma-ray astronomy can be found in several recent reviews \cite{Fegan 2012}, \cite{Lorenz 2012}, \cite{Hillas 2013}.

\section{Outline of the review}
With the present introduction, the first volume of this review includes two articles on instrumentation (from space and from the ground), a theoretical basis on particle acceleration in astrophysical plasmas and on radiation mechanisms, and articles on several galactic objects: pulsars, binary systems, supernova remnants and pulsar wind nebulae, and the Galactic Centre region.
The second volume covers extragalactic sources (starburst galaxies, active galactic nuclei and gamma-ray bursts) 
and applications to fundamental physics (search for primary black holes and Dark Matter, 
search for Lorentz invariance violation and for axion-like particles). 
The second volume is completed by a review of future projects in $\gamma$-ray astronomy. 
\section{Glossary}
The acronyms currently used in
the articles of the two volumes are explicited in the following glossary.\\

\begin{description}
\item[1FGL,2FGL,3FGL]: first, second and third catalogs of Fermi LAT.
\item[1LAC,2LAC,3LAC]: first, second and third AGN catalogs of Fermi LAT.
\item[ACT]: Atmospheric Cherenkov Telescope.
\item[AdEPT]: Advanced Energetic Pair Telescope.
\item[AGILE]: Astro-rivelatore Gamma a Immagini LEggero.
\item[AGN]: Active Galactic Nucleus.
\item[ALP]: Axion-like particle.
\item[ASGAT]: AStronomie GAmma \`a Th\'emis.
\item[AU]: Astronomical unit.
\item[BATSE]: Burst and Transient Source Experiment (on board CGRO satellite).
\item[BH]: Black hole.
\item[BL Lac]: Class of active galactic nuclei similar to BL Lacertae.
\item[BW]: Black widow.
\item[CALET]: CALorimetric Electron Telescope.
\item[CANGAROO]: Collaboration of Australia and Nippon for a GAmma Ray Observatory in the Outback.
\item[CAT]: Cherenkov Array at Th\'emis.
\item[CELESTE]: CErenkov Low Energy Sampling and Timing Experiment.
\item[CGRO]: Compton Gamma-Ray Observatory.
\item[CL]: Confidence limit.
\item[CMB]: Cosmic Microwave Background.
\item[CMZ]: Central molecular zone (part of the Galactic Centre region)
\item[COMPTEL]: Compton telescope (on board CGRO satellite).
\item[CR]: Cosmic Rays.
\item[CTA]: Cherenkov Telescope Array.
\item[CWB]: Colliding wind binary.
\item[DAMPE]: Dark Matter Particle Explorer (satellite).
\item[DIGB]: Diffuse isotropic gamma-ray background.
\item[DM]: Dark matter.
\item[DSA]: Diffusive shock acceleration.
\item[DSSD]: Double-sided silicon strip detector.
\item[EAS]: Extensive air shower.
\item[EBL]: Extragalactic background light.
\item[EED]: Electron energy distribution.
\item[EGB]: Extragalactic gamma-ray background.
\item[EGRET]: Energetic Gamma Ray Experiment Telescope  (on board CGRO satellite).
%\item[EVPA]: Electric vector polarization angle.
\item[FIDO]: Force-free Inside the light cylinder and Dissipative Outside (model of pulsar environment).
\item[FR~I, FR~II]: Fanaroff-Riley I and II (radiogalaxies).
\item[FSRQ]: Flat spectrum radio-quasar.
\item[GBM]: Gamma Ray Burst Monitor (on board Fermi Gamma-ray Space Telescope).
\item[GAMMA-400]: Gamma Astronomical Multifunctional Modular Apparatus with the
maximum gamma-ray energy of 400 GeV.
\item[GC]: Galactic Centre.
\item[GJ]: Goldreich-Julian (model of pulsar magnetospheres).
\item[GLAST]: Gamma-ray Large Area Space Telescope (former name of the Fermi space
telescope).
\item[GRAAL]: Gamma Ray Astronomy at ALmeria.
\item[GRB]: Gamma Ray Burst.
\item[HAGAR]: High Altitude GAmma Ray (telescope).
\item[HARPO]: Hermetic ARgon POlarimeter.
\item[HAWK]: High Altitude Water Cherenkov (telescope).
\item[HEGRA]: High Energy Gamma Ray Astronomy (array of 5 telescopes).
\item[HE]: High energy (100 MeV-100 GeV).
\item[HERD]: High Energy cosmic Radiation Detection.
\item[H.E.S.S.]: High Energy Stereoscopic System.
\item[HiSCORE]: Hundred Square-km Cosmic ORigin Explorer.
\item[HSP]: High-synchrotron-peak (blazar).
\item[HVC]: High velocity cloud.
\item[INTEGRAL]: INternational Gamma-Ray Astrophysics Laboratory.
\item[ISM]: Interstellar medium.
\item[ISP]: Intermediate-synchrotron-peak (blazar).
\item[IACT]: Imaging Atmospheric Cherenkov Telescope.
\item[IC]: Inverse Compton radiation.
\item[ICRC]: International Cosmic Ray Conference.
\item[KM3Net]: Cubic Kilometer NEutrino Telescope.
\item[LAT]: Large Area Telescope (Fermi Gamma ray Space Telescope).
\item[LHAASO]: Large High Altitude Air Shower Observatory.
\item[LMC]: Large Magellanic Cloud.
\item[LMXB]: Low-mass X-ray binary.
\item[LIV]: Lorentz invariance violation.
\item[LSP]: Low-synchrotron peak (blazar).
\item[LST]: Large size telescopes (in CTA).
\item[MACE]: Major Atmospheric Cherenkov Experiment.
\item[MAGIC]: Major Atmospheric Gamma-ray Imaging Cherenkov telescope.
\item[MAGN]: Misaligned active galactic nuclei.
\item[MHD]: Magneto hydrodynamics.
\item[MSP]: Millisecond pulsar.
\item[MST]: Medium size telescopes (in CTA).
\item[NFW]: Navarro-Frenk-White (halo profile).
\item[NIR]: Near infrared
\item[OSO-3]: Third satellite of the Orbiting Solar Observatory program.
\item[OSSE]: Oriented Scintillation Spectrometer Experiment (on board CGRO
satellite).
\item[PAM]: Photon-ALP mixing.
\item[PANGU]: PAir-productioN Gamma-ray Unit.
\item[PBH]: Primordial black hole.
\item[PIC]: Particle-in-cell (simulations).
\item[PMT]: Photo-multiplier tube.
\item[PWN]: Pulsar wind nebula.
\item[QCD]: Quantum chromodynamics.
\item[QED]: Quantum electrodynamics.
\item[QG]: Quantum gravity.
\item[RB ]: Redback.
\item[RL-NLS1]: Radio-loud narrow-line Seyfert 1 (galaxy).
\item[SAS-2]: second Small Imaging Satellite.
\item[SSC]: Synchrotron Self Compton (model).
\item[SSD]: Silicon strip detector.
\item[SED]: Spectral Energy Distribution.
\item[SEP]: Strong equivalence principle.
\item[SMBH]: Supermassive black hole.
\item[SN]: Supernova.
\item[SNR]: Supernova remnant.
\item[SR]: Synchrotron radiation.
\item[SST]: Small size telescopes (in CTA).
\item[STACEE]: Solar Tower Atmospheric Cherenkov Effect Experiment.
\item[SVOM]: Space-based multi-band astronomical Variable Objects Monitor.
\item[SW]: Striped wind.
\item[THEMISTOCLE]: Tracking High Energy Muons In Showers Triggered On Cherenkov Light Emission.
\item[UHE]: Ultra-high energy (above 1 PeV).
\item[VERITAS]: Very Energetic Radiation Imaging Telescope Array System.
\item[VHE]: Very high energy (above 100 GeV).
\item[VLBI]: Very-long-base interferometry.
\item[WCD]: Water Cherenkov detectors.
\item[WIMP]: Weakly interacting massive particle.
\item[WMAP]: Wilkinson Microwave Anisotropy Probe.
\end{description}
\section{Acknowlegements}
We are grateful to Dr.~Steve Fegan for fruitful discussions in the preparation of the article.
Figures \ref{fig:sed}, \ref{fig:oso-cosb} and \ref{fig:egret_fermi} are reproduced by permission of AAS. 
%%%%%%%%%%%%%%%%%%%%%%
% Insert here Fig. 6
% Figure 6.  Sky map of the present 155 VHE gamma-ray sources (TeVCat \cite{TeVCat}) in galactic coordinates (Hammer projection) with a zoom on the Galactic Centre area.]
%%%%%%%%%%%%%%%%%%%%%%
%%\begin{center}
%\includegraphics[width=0.45\linewidth]{vhe_skymap.eps}
%\caption{\it Sky map of the present 155 VHE gamma-ray sources (TeVCat \cite{TeVCat}) in galactic coordinates (Hammer projection)
%with a zoom on the Galactic Centre area.}
%\label{fig:vhe_skymap}
%\end{center}
%\end{figure}

%%%%%%%%%%%%%%%%%%%%%%%%%%%%%%%%%%%%%%%%%%%%%%%%%%%%%%%%%

\end{document}